\documentclass[fleqn,usenatbib]{mnras}

\usepackage{newtxtext,newtxmath}
\usepackage[T1]{fontenc}
\usepackage{graphicx}	% Including figure files
\usepackage{amsmath}	% Advanced maths commands
\usepackage{subcaption}
\usepackage{xcolor}
\usepackage{soul}       % Highlight text
\usepackage{scalerel}

\def\rhalf{\mbox{$r_{1/2}$}}

\def\Msun{\, M_{\odot}}

\def\Rtwohc{R_{\rm 200c}}
\def\Mtwohc{M_{\rm 200c}}
\def\Mtot{M_{\rm tot}}
\def\Ms{M_{\star}}
\def\rhalf{r_{1/2}}
\def\Rhalf{R_{1/2}}

\def\Mhalf{\mathcal{M}_{\rm tot}(<\rhalf)}

\def\rhotot{\rho_{\rm tot}}
\def\rhohalf{\rho_{\rm tot}(<\rhalf)}

\title[Densities and assembly histories of $\Lambda$CDM satellites]{Densities and mass assembly histories of the Milky Way satellites\\ are not a challenge to $\Lambda$CDM}

\author[Kravtsov \& Wu]{
Andrey Kravtsov$^{1,2,3}\thanks{E-mail: kravtsov@uchicago.edu}$ and Zewei Wu$^1\thanks{E-mail: wuz25@uchicago.edu}$
\\
$^{1}$Department of Astronomy  \& Astrophysics, The University of Chicago, Chicago, IL 60637 USA\\
$^{2}$Kavli Institute for Cosmological Physics, The University of Chicago, Chicago, IL 60637 USA\\
$^{3}$Enrico Fermi Institute, The University of Chicago, Chicago, IL 60637 USA
}

\date{Accepted XXX. Received YYY; in original form ZZZ}

\pubyear{2023}

\begin{document}
\label{firstpage}
\pagerange{\pageref{firstpage}--\pageref{lastpage}}
\maketitle

\begin{abstract}
We use the \texttt{GRUMPY} galaxy formation model based on a suite of zoom-in, high-resolution, dissipationless $\Lambda$CDM simulations of the Milky Way (MW) sized haloes to examine total matter density within the half-mass radius of stellar distribution, $\rho_{\rm tot}(<\rhalf)$, of satellite dwarf galaxies around the MW hosts and their mass assembly histories. We compare
model results to $\rho_{\rm tot}(<\rhalf)$ estimates for observed dwarf satellites of the Milky Way spanning their entire luminosity range. We 
show that observed MW dwarf satellites exhibit a trend of decreasing total matter density within a half-mass radius, $\rhohalf$, with increasing stellar mass. This trend is in general agreement with the trend predicted by the model. None of the observed satellites are overly dense compared to the results of our $\Lambda$CDM-based model.  We also show that although the halo mass of many satellite galaxies is comparable to the halo mass of the MW progenitor at $z\gtrsim 10$, at these early epochs halos that survive as satellites to $z=0$ are located many virial radii away from the MW progenitors and thus do not have a chance to merge with it. Our results show that neither the densities estimated in observed Milky Way satellites nor their mass assembly histories pose a challenge to the $\Lambda$CDM model.  In fact, the broad agreement between density trends with the stellar mass of the observed and model galaxies can be considered as yet another success of the model. 
\end{abstract}


%%%%%%%%%%%%%%%%%%%%%%%%%%%%%%%%%%%%%%%%%%%%%%%%%%

%%%%%%%%%%%%%%%%% BODY OF PAPER %%%%%%%%%%%%%%%%%%

%----------------------
\section{Introduction}
\label{sec:intro}
%----------------------

Dwarf galaxies orbiting around the Milky Way allow us to study galaxy formation and test $\Lambda$ Cold Dark Matter model (hereafter $\Lambda$CDM) at the smallest scales \citep[see, e.g.,][for reviews]{Read.etal.2016,bullock_boylankolchin17,Sales.etal.2022}. 
In particular, total matter density estimates in the inner regions of dwarf galaxies are used as one of the key tests of $\Lambda$CDM model \citep[e.g.,][]{Boylan_Kolchin.etal.2011,Oh.etal.2015,Read.etal.2016} and can be used to distinguish between models in which dark matter has different degrees of self-interaction \citep[e.g.,][]{Silverman.etal.2023}.

\citet{safarzadeh_loeb21} recently used density within half-light radius estimated for a sample of observed nearby dwarf galaxies and some additional model assumptions to estimate the formation epoch of their parent haloes and their mass at that epoch. They argued that some of the dwarf galaxies with the largest densities have masses at $z\sim 3-5$ that are comparable to the expected mass of the Milky Way (hereafter MW) progenitor and that this fact is a challenge to the $\Lambda$CDM model. 

In this study, we examine this challenge in detail. We use a suite of high-resolution Caterpillar simulations of MW-sized halos \citep{Griffen.etal.2016} and a model of dwarf galaxy formation of \citet{Kravtsov.Manwadkar.2022} to model the evolution of the MW satellites and their observable properties, such as $V$-band luminosities and half-mass radii, $r_{1/2}$. The model predicts the total mass density within $r_{1/2}$, $\rho_{\rm tot}(<r_{1/2})$, and we compare the predicted densities with the estimates for observed MW satellites. For the latter, we use an up-to-date compilation of half-light radii, luminosities, and stellar velocity measurements for the entire range of stellar masses of the observed MW satellites
and the dynamical mass estimator of \citet{Wolf.etal.2010}.  

We find that $\rho_{\rm tot}(<r_{1/2})$ of observed satellites decreases with increasing stellar mass and this relation is reproduced by the model. There is thus no discrepancy between $\rho_{\rm tot}(<r_{1/2})$  of the MW satellites and predicted densities of satellites in the $\Lambda$CDM model. 
Likewise, we find that the masses of some of the surviving MW satellites have likely been comparable to the MW mass at $z\gtrsim 10$. However, this is not an issue because at that time these satellites were far from the main MW progenitor and this is why they did not merge with it. 

We describe the simulations used and the modelling of the luminous Milky Way satellite systems in Section~\ref{sec:modelling} and describe observational data used in our comparisons in Section~\ref{sec:data}. We present our main results in Section~\ref{sec:results} and summarize our results and conclusions in Section~\ref{sec:summary}.

%-------------------------------
\section{Modelling Milky Way satellite system}
\label{sec:modelling}
%-------------------------------

We model the population of Milky Way dwarf satellite galaxies around the Milky Way using tracks of haloes and subhaloes from the Caterpillar \citep{Griffen.etal.2016} suite of $N$-body simulations\footnote{\url{https://www.caterpillarproject.org}} of 32 MW-sized haloes.  We use the highest resolution suite LX14 to maximize the dynamic range of halo masses probed by our modelling. 
 
The haloes were identified using the modified version of the Rockstar halo finder and the Consistent Trees Code \citep{Behroozi.etal.2013}, with modification improving recovery of the subhaloes with high fraction of unbound particles \citep[see discussion in Section 2.5 of][]{Griffen.etal.2016}. As was shown in \citet{Manwadkar.Kravtsov.2022} (see their Fig. 1), the subhalo peak mass function in the LX14 simulations is complete at $\mu=M_{\rm peak}/M_{\rm host} \gtrsim 4 \times 10^{-6}$, or $M_{\rm peak} \approx 4 \times 10^{6} \Msun$ for the host halo mass $M_{\rm host}\approx 10^{12}\, M_\odot$, even in the innermost regions of the host ($r < 50$ kpc).  This is sufficient to model the full range of luminosities of observed Milky Way satellites, as faintest ultrafaint dwarfs are hosted in haloes of $M_{\rm peak}\gtrsim 10^7\, M_\odot$ in our model \citep[][]{Kravtsov.Manwadkar.2022,Manwadkar.Kravtsov.2022}. Moreover, in this study we use only galaxies hosted in haloes with $M_{\rm peak}>10^8\, M_\odot$. 

The mass evolution tracks of subhaloes of MW-sized host haloes are used as input for the \texttt{GRUMPY} galaxy formation model, which evolves various properties of gas and stars of the galaxies they host. As a regulator-type galaxy formation framework \citep[e.g.,][]{Krumholz.Dekel.2012,Lilly.etal.2013,Feldmann.2013}, \texttt{GRUMPY} is designed to model galaxies of $\lesssim L_\star$ luminosity \citep[][]{Kravtsov.Manwadkar.2022}, which follows the evolution of a number of key galaxy properties by solving a system of coupled differential equations. The model accounts for UV heating after reionization and associated gas accretion suppression onto small mass haloes, galactic outflows, a model for gaseous disk and its size, molecular hydrogen mass, star formation, etc.  The evolution of the half-mass radius of the stellar distribution is also modelled. The galaxy model parameters used in this study are identical to those used in \citet{Manwadkar.Kravtsov.2022}. 

Here we use the model to predict luminosities, stellar masses, and stellar half-mass radii (sizes) of satellite galaxies around the MW-sized haloes from the Caterpillar suite.  To estimate luminosity in the $V$-band filter we use the Flexible Stellar Population Synthesis (\texttt{FSPS}) code \citep{Conroy.etal.2009,Conroy.etal.2010} in conjunction with its Python bindings, \textsc{PyFSPS} \footnote{\href{https://github.com/dfm/python-fsps}{\tt https://github.com/dfm/python-fsps}} and star formation histories and metallicity evolution of the model galaxies.

The \texttt{GRUMPY} model is described and tested against a wide range of observations of local dwarf galaxies in \citet[][]{Kravtsov.Manwadkar.2022}. Importantly for this study, the model was shown to reproduce luminosity function and radial distribution of the Milky Way satellites and size-luminosity relation of observed dwarf galaxies \citep{Manwadkar.Kravtsov.2022}. We thus use luminosities of model galaxies to select counterparts of observed MW satellites, while half-mass radii are used to estimate total mass and density within these radii, as we describe in the next section.  

The cosmological parameters adopted in this study are those of the Caterpillar simulation suite: $h=H_0/100=0.6711$, $\Omega_{\rm m0}=0.32$, $\Omega_\Lambda=0.68$. 

%---------------------------------------------------------
\subsection{Estimating densities of model galaxies}
\label{sec:rhohalf_model}
%------------------------------------

To estimate total matter densities $\rhotot(<\rhalf)$ for model galaxies, we use individual half-mass radii $\rhalf$ predicted for each model galaxy by the \texttt{GRUMPY} model:
\begin{equation}
    \rhohalf = \frac{3 M_{\rm tot}(<\rhalf)}{4\pi\rhalf^3}
    \label{eq:rhohalf}
\end{equation}
To estimate $M_{\rm tot}(<\rhalf)$ we consider three assumptions for the density profile of dark matter. 

In the first case, we adopt the Navarro-Frenk-White density profile \citep[NFW,][]{Navarro.etal.1997} for dark matter profile and use $\Mtwohc$, $\Rtwohc$, and the NFW scale radius, $r_s$, available in the halo tracks for a grid of cosmic time $t_i$, adding half of the current stellar mass predicted by the model: 
\begin{equation}
    M_{\rm tot,NFW}(<\rhalf) = M_{\rm dm,NFW}(<\rhalf) + \frac{1}{2}\,\Ms
\label{eq:mhalfnfw}
\end{equation}
where
\begin{equation}
    M_{\rm dm,NFW}(<r) = \Mtwohc\,\frac{f(r/r_s)}{f(\Rtwohc/r_s)};
\label{eq:mdmnfw}
\end{equation}
and
\begin{equation}   
    f(x) = \ln(1+x) - \frac{x}{x+1}.
\end{equation}
In the equation~\ref{eq:mhalfnfw} above we add only stellar mass, assuming that for satellites all of the gas mass is stripped, as is the case for the observed Milky Way satellites with exception of the Large and Small Magellanic Clouds (LMC and SMC hereafter). The contribution of stars and gas to the total mass within $\rhalf$ is quite small and this assumption has negligible effect on our results. 

In the second case, we adopt the dark matter density profile modified by stellar feedback effects proposed by \citet{Read.etal.2016}: 
\begin{equation}
    M_{\rm tot, R}(<\rhalf) = M_{\rm dm,Read}(<\rhalf) + \frac{1}{2}\,\Ms,
\label{eq:mhalfread}
\end{equation}
where $M_{\rm dm,Read}(<\rhalf)$ is the feedback modified mass within $\rhalf$ computed using equations in Section 3.3.3 of \citet{Read.etal.2016}. These equations
depend on the duration of star formation parameter $t_{\rm sf}$ and scale radius of the galaxy halo $r_s$: we adopt the difference between the time where a model galaxy formed $90\%$ of its stellar mass and the  start of the galaxy evolution track as $t_{\rm sf}$ and use individual $r_s$ from the $z=0$ halo catalog. On average, for most dwarf galaxies in the MW satellite mass range, the \citet{Read.etal.2016} model leads to reduction of mass within $\rhalf$ by a factor of two, even in the faintest galaxies. This is in line with the finding by these authors that core in the dark matter distribution forms in haloes of all masses, as long as star formation proceeds sufficiently long. 

In the third case, we adopt the dark matter density profile of \citet{Lazar.etal.2020}, which approximates effects of stellar feedback on the density profile in the FIRE-2 simulations: 
\begin{equation}
    M_{\rm tot, L}(<\rhalf) = M_{\rm dm,Lazar}(<\rhalf) + \frac{1}{2}\,\Ms
\label{eq:mhalflazar}
\end{equation}

Specifically, we use parametrization of the cored-Einasto density profile in the equations~8-10, 12 of \citet{Lazar.etal.2020} and equations for the cumulative mass profile in their Appendix B1 and parameters in the second row of their Table 1 for the dependence of profile as a function of stellar mass $M_\star$. We chose dependence on the stellar mass, to minimize effects of different $M_\star/M_{\rm h}$ in their simulations and in our model. Note that the profiles were calibrated only for galaxies of $M_\star\gtrsim 10^5\,M_\odot$. However, in this model effects of feedback for $M_\star<10^6\, M_\odot$ are expected to be negligible and thus extrapolating their results to smaller masses is equivalent to simply assuming Einasto profile with negligible core for these low-mass systems. 

%------------------
\section{Observed dwarf galaxy measurements}
\label{sec:data}
%------------------

We use a sample of observed MW dwarf satellites and their $V$-band luminosities, projected half-light radii $R_{1/2}$, 
and velocity dispersions compiled from the literature, with some updates and modifications to make estimates of some of the absolute magnitudes and sizes more uniform. The sample and its compilation is described  in the Appendix~\ref{app:sample}.

%------------------
\subsection{Estimating masses and densities for observed dwarf satellites}
\label{sec:rhohalf_obs}
%------------------
To estimate masses $\Mtot(<\rhalf)$  using estimator given by equation 2 of \citet{Wolf.etal.2010}:
\begin{equation}
    \Mhalf = 930\,\sigma_{\star,\rm los}^2\,\Rhalf\ M_\odot, 
    \label{eq:mhalf}
\end{equation}
where $\sigma_{\star,\rm los}$ is the line of sight velocity dispersion of stars in km/s, $\Rhalf$ is projected half-light radius in parsecs, and $\rhalf$ is the 3D stellar half-mass radius. 

This type of estimator is known to be robust for spheroidal systems as it is not sensitive to the velocity anisotropy of the stellar motions \citep{Walker.etal.2009,Churazov.etal.2010,Errani.etal.2018} and to differences in the density profile \citep{Amorisco.Evans.2011}. Nevertheless, the estimator may be somewhat biased \citep[e.g.,][although see \citealt{Gonzalez_Samaniego.etal.2017}]{Campbell.etal.2017,Errani.etal.2018}. The magnitude of the bias is small in stellar systems that are velocity dispersion-dominated and larger in rotation-dominated systems, but even for the latter, the bias is typically less than a factor of two which is smaller than a typical error in observational estimates of $\Mhalf$. It is also smaller than the scatter of 
$\Mhalf$ values expected for model galaxies at a given stellar mass.

Given $\Mhalf$ we estimate densities $\rhotot(<\rhalf)$ for observed dwarf satellites using equation~\ref{eq:rhohalf}. 
This requires conversion of the projected half-light radii $\Rhalf$ estimated from observations to the 3d half-mass radii $\rhalf$. 
This conversion, however, is somewhat uncertain because it depends on the star formation history of galaxies \citep{Somerville.etal.2018,Suess.etal.2019} and ellipticity and radial density distribution of stars \citep{Somerville.etal.2018,Behroozi.etal.2022}. The factor $\chi$ relating the two radii $\rhalf=\chi \Rhalf$ is thus expected to vary between $\chi\approx 0.85-1$ for disky systems to $\chi\approx 1.34-1.6$ for spheroidal systems. 

Given these dependencies, such conversion would need to be done carefully for individual galaxies, given that observed ellipticities of Milky Way satellites vary fairly widely. However, information to estimate $\chi$ reliably is lacking for many of the galaxies. We therefore 
chose to keep $\chi=1$ for this analysis, but note that for most galaxies in the sample we expect $\chi$ in the range $\approx 1-1.5$ and that their densities $\rhotot(<\rhalf)$ thus may be somewhat {\it overestimated} in the figures below. This makes our conclusions that observed dwarf satellites are not overly dense compared to $\Lambda$CDM expectation even stronger.   
On the other hand, this conversion uncertainty does not affect the estimate of $\Mtot(<\rhalf)$ because it is expected to estimate 
total mass within a 3d half-light radius.

%------------------
\section{Results}
\label{sec:results}
%------------------

%------------------------------------------------------------------------------------
\subsection{Comparing $M(<r_{1/2})-L_V$ relations in the model and observed galaxies}
\label{sec:mhalf_lv_comp}
%-------------------------------------------------------------------------------------

\begin{figure}
   \centering
   {\includegraphics[width=0.49\textwidth]{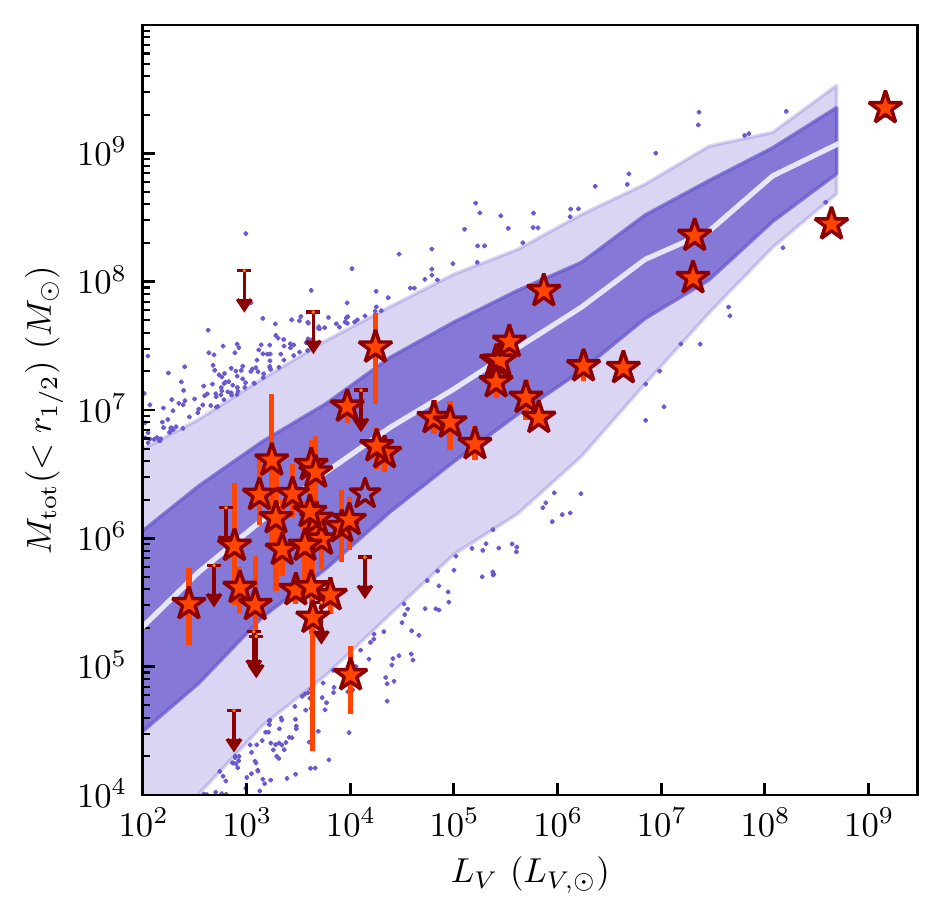}}
   \caption{The total mass within half-mass radius, $M_{\rm tot}(<\rhalf)$ vs $V$-band galaxy luminosity for the observed MW satellites ({\it red stars and arrows}) and model dwarf satellites (white line, shaded bands and blue dots) estimated as described in Section~\ref{sec:rhohalf_obs} and ~\ref{sec:rhohalf_model}, respectively. The downward arrows in the observed sample are galaxies for which only an upper limit on the velocity dispersion (and hence on the mass) exists currently. The white solid line shows the median $M_{\rm tot}(<\rhalf)-L_V$ relation for model galaxies within virial radii of the MW-sized haloes in the suite, while dark and light shaded blue bands show the 68\% and 95\% of the distribution of model galaxies around the median. Model galaxies outside the shaded bands are shown by the blue dots. The figure shows that the model reproduces $M_{\rm tot}(<\rhalf)-L_V$ relation of observed galaxies, which indicates the model forms galaxies of a given luminosity in haloes of correct mass.}
   \label{fig:mhalf_lv}
\end{figure}

Observed dwarf satellites of the Milky Way exhibit a correlation of the total mass within half-mass radius, $M_{\rm tot}(<\rhalf)$, and their luminosity \citep[e.g., see Fig.~4 in][]{Simon.2019}. $M_{\rm tot}(<\rhalf)-L_V$ relations of the observed and model galaxies are compared in Figure~\ref{fig:mhalf_lv}. 

The figure shows that our model reproduces both the normalization and the form of the observed correlation. The median relation shown as a solid line in Figure~\ref{fig:mhalf_lv} can be accurately described by the following power law over the entire mass range shown: 
\begin{equation}
M_{\rm tot}(<\rhalf)\approx 10^6\,M_\odot\left(\frac{L_V}{10^3 L_{V,\odot}}\right)^{0.55}. 
\label{eq:lv_mhalf_corr}
\end{equation}
This relation reflects relation of $\rhalf$ and parent halo virial radius and the relation between luminosity and halo mass, as discussed in more detail in the Appendix~\ref{app:lv_mhalf}. Note that the scatter in the model relation is much larger than the expected scatter of the halo mass within a fixed radius. 
This is because the scatter in the $\rhalf-M_{\rm 200c}$ relation is substantial due to scatter in both $L_V-M_{\rm 200c}$ and $\rhalf-L_V$ relations.

Agreement between observed and model correlation indicates the model galaxies of a given luminosity form in haloes of correct mass and with $\rhalf$ values consistent with those of observed galaxies \citep[see also Fig. 12 for an explicit comparison of $\rhalf-M_V$ relations of the galaxies in our model and observed dwarf galaxies][]{Manwadkar.Kravtsov.2022}. We can therefore meaningfully consider both densities $\rho_{\rm tot}(<\rhalf)$ and mass assembly histories of the model galaxies to examine the ostensible challenge to $\Lambda$CDM.

\begin{figure*}
   \centering {\includegraphics[width=0.49\textwidth]{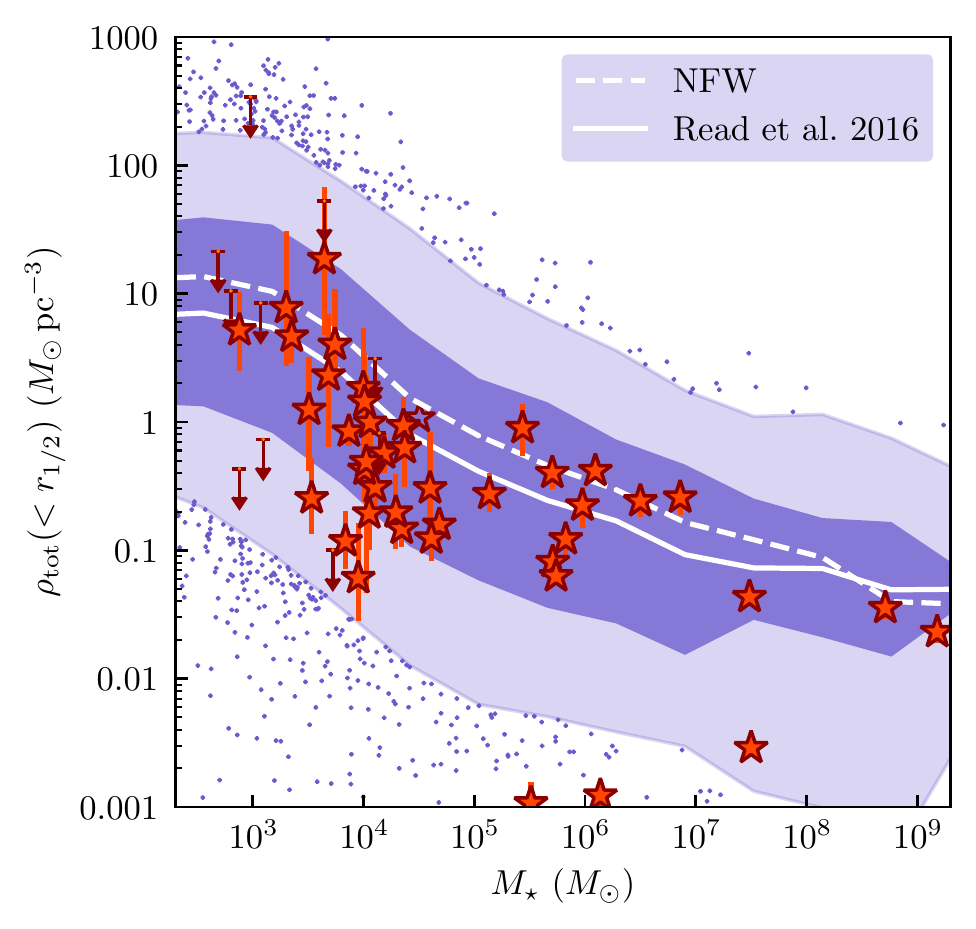}
   \includegraphics[width=0.49\textwidth]{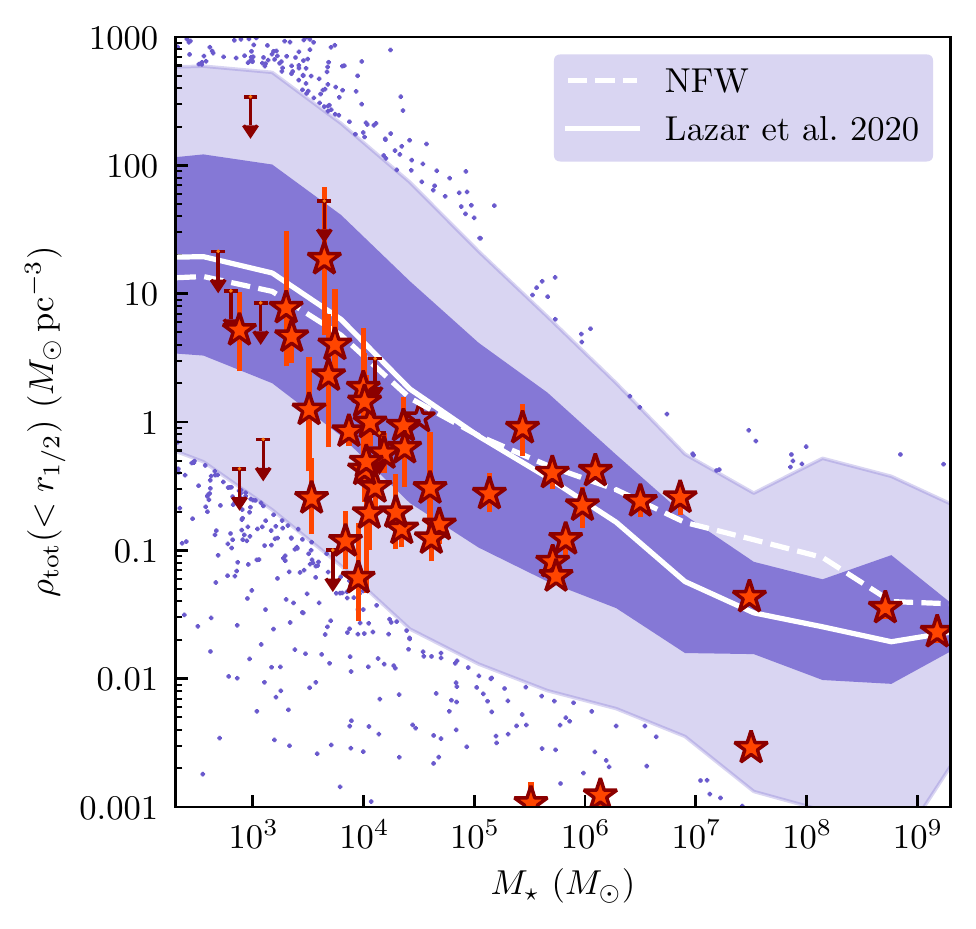}}
   \caption{Total density $\rhohalf$ within $\rhalf$, where dark matter mass within $\rhalf$ is assumed to follow the feedback-modified profile of \citet[][left panel]{Read.etal.2016} and \citet[][right panel]{Lazar.etal.2020}, as described in Section~\ref{sec:rhohalf_model}. In both panels the dashed lines show the median $\rho_{\rm tot}(<\rhalf)$ when NFW profile is assumed instead to compute $M_{\rm dm, NFW}(\rhalf)$. 
   The shaded areas show the $68.27\%$ and $95.45\%$ distribution for model galaxies in a given stellar mass bin around the medians shown by solid lines.
   Individual galaxies outside the $95.45\%$ band are shown by points. In both panels estimates for observed dwarf MW satellites are shown by red stars and upper limits (see Section~\ref{sec:rhohalf_obs}): the open star shows Centaurus I for which $\sigma_{\star,\rm los}$ is reported by \citet{Martinez_Vazquez.etal.2021} without uncertainty. The downward arrows show upper limits on density for galaxies for which only the upper limit on velocity dispersion is obtained so far.}
   \label{fig:rhohalf_ms}
\end{figure*}

%------------------------------------
\subsection{Densities of the Milky Way dwarf satellites}
\label{sec:densities}
%------------------------------------

Figure~\ref{fig:rhohalf_ms} shows $\rho_{\rm tot}(<\rhalf)$, total mass 
density within $\rhalf$, for the model and observed dwarf satellite galaxies located within virial radius of each MW-sized halo in the Caterpillar suite. The two panels show the same $\rho_{\rm tot}(<\rhalf)$ measurements for observed satellites (see Section~\ref{sec:rhohalf_obs}) by red stars and arrows showing upper limits, while for the model galaxies the mass is computed using parameters of the parent subhalo and model galaxy, but  assuming the feedback-modified profiles of \citet{Read.etal.2016} in the left panel and of \citet{Lazar.etal.2020} in the right panel (see Section~\ref{sec:rhohalf_model}). The median relation in the case when the NFW density profile (not modified by feedback) is assumed instead is shown by the dashed line in each panel.  

The effects of feedback expected to  modify dark matter density profiles for larger dwarf galaxies are uncertain in the ultra-faint dwarf regime. Several studies  indicate that feedback effects should be confined to the galaxies with $M_\star/M_{\rm halo}\sim 10^{-4}-10^{-1}$ \citep{DiCintio.etal.2021,Tollet.etal.2016,Lazar.etal.2020}. However, results of \citet{Read.etal.2016} indicate that significant modification of the central density distribution occurs in halos of all relevant masses, as long as galaxy is able to form stars for a sufficiently long time.

Regardless of the assumptions about dark matter density profile the model broadly reproduces the overall trend 
exhibited by observed galaxies, although observed ultra-faint galaxies ($M_\star\lesssim 10^5\, M_\odot$) tend to have somewhat lower densities than the model ones. We have checked that this is also the case if we only select model dwarf galaxies within the central 70 kpc. 
The two observed outliers at low density of $\rhohalf\approx 0.001\, M_\odot\,\rm pc^{-3}$ are Crater II and Antlia II galaxies, which may be undergoing tidal disruption by the Milky Way \citep[][although see \citealt{Borukhovetskaya.etal.2022}]{Ji.etal.2021,Vivas.etal.2022,Pace.etal.2022}

It is interesting to note that aside from these reported outliers, there is no apparent ``diversity problem'' -- or surprisingly large scatter -- in the distribution of $\rhohalf$ for observed dwarf satellites compared to the model results. Such diversity of mass profiles exists for larger dwarf galaxies, where the mass profile is estimated using their observed rotation curves \citep{Oman.etal.2015}. This may partly be due to the large scatter in the subhalo profiles compared to their isolated counterparts due to the varying amounts of tidal stripping that they experience.

This overall trend of decreasing density with increasing stellar mass is expected in $\Lambda$CDM because 1) galaxies of larger $M_\star$ form in halos of larger halo mass $M_{\rm h}$, on average, and 2) $\rhalf$ is roughly a fixed fraction of the virial radius \citep[e.g.,][]{Kravtsov.2013}. At a fixed fraction of the virial radius, smaller mass halos are predicted to be denser in the CDM scenario. In fact, 
in our model $\rhalf\approx 0.03R_{\rm 200c}$ for galaxies with $M_\star\gtrsim 10^6\, M_\odot$, but the proportionality factor drops to $\rhalf\approx 0.005R_{\rm 200c}$ for the faintest $M_\star\sim 10^3\,M_\odot$ galaxies. This additional decrease results in a faster increase of $\rhohalf$ towards smaller masses than expected for the CDM haloes at a fixed fraction of the virial radius.

\begin{figure*}
   \centering
   {\includegraphics[width=0.33\textwidth]{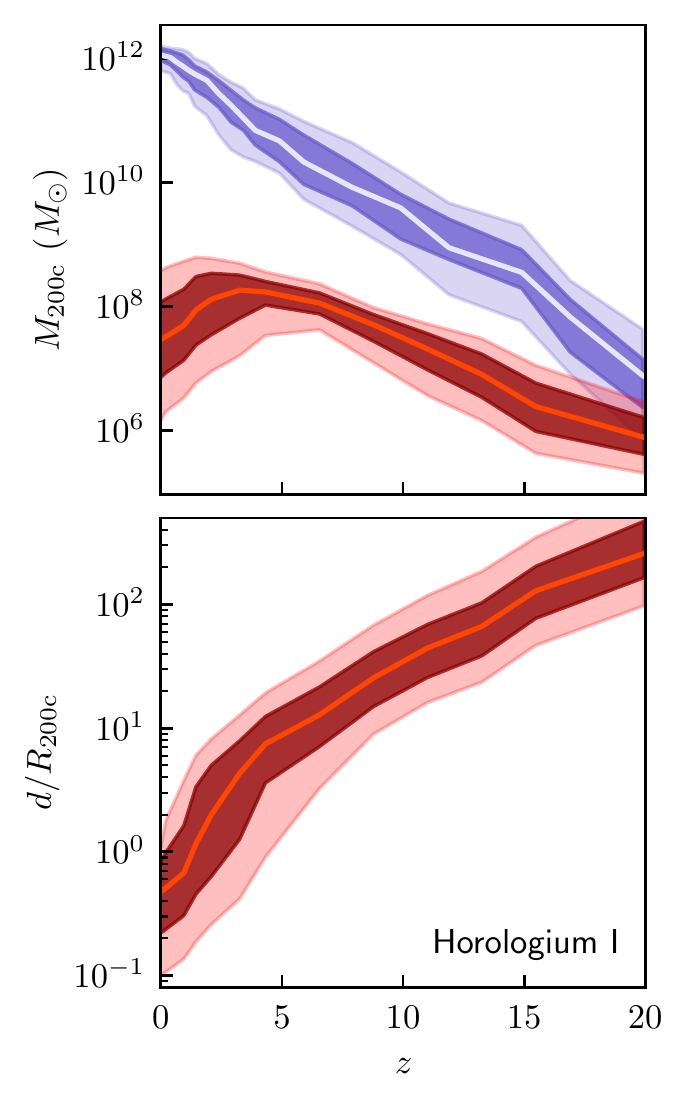}
   \includegraphics[width=0.33\textwidth]{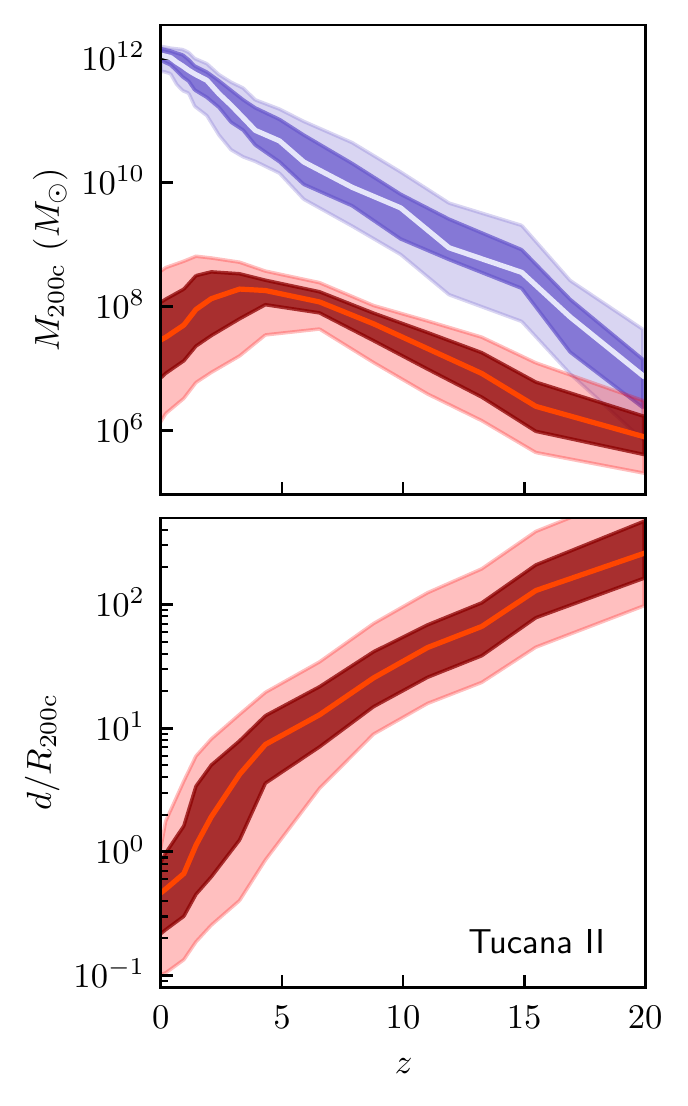}
   \includegraphics[width=0.33\textwidth]{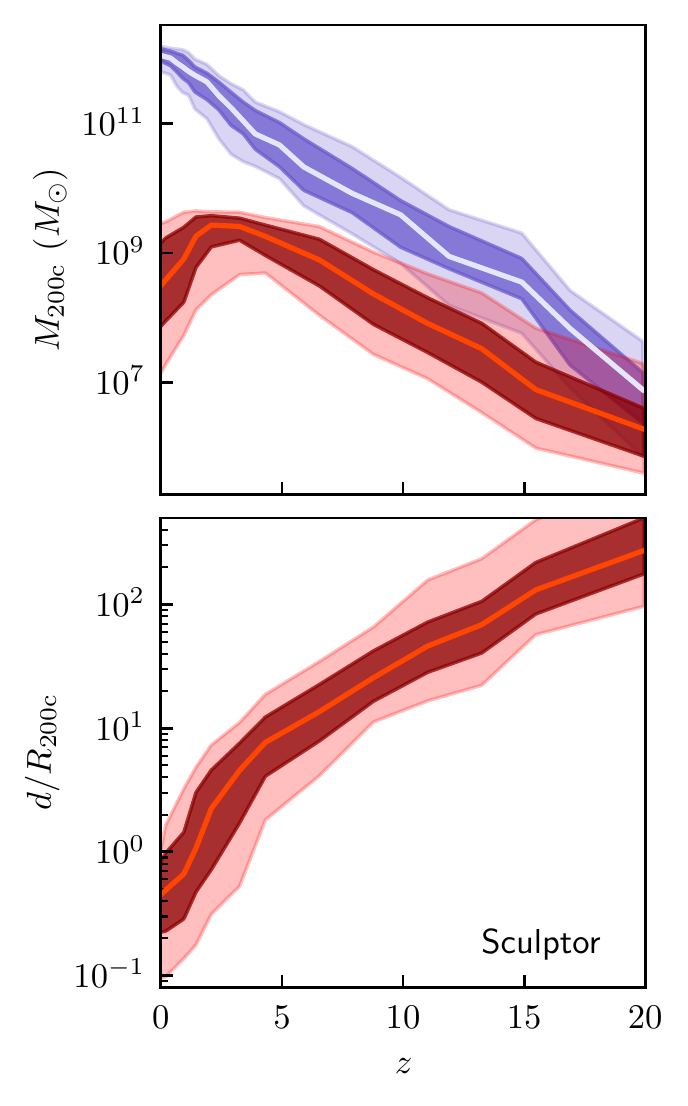}}
   \caption{Upper panel: the redshift halo mass evolution of the main progenitors of the MW-sized hosts and of the subhalos that host galaxies at $z=0$  with absolute $V$-band magnitudes within $\pm 0.5$ of the $M_V$ value of the three representative MW satellite galaxies labeled in the lower right corner of each column.  Lower panel: evolution of the distance between satellite and MW progenitor halos in units of the MW progenitor virial radius $R_{\rm 200c}(z)$.   Evolution shown in these three panels is typical of all satellite galaxies in the simulations.  }
   \label{fig:mah_revo}
\end{figure*}

None of the observed galaxies has a surprisingly high density within $\rhalf$ compared to model expectations. We note that the updated values of velocity dispersion and half-light radius we use for the Horologuium I and Tucana II result in the $\rhohalf$ of $4.0\, M_\odot\,\rm pc^{-3}$ and $0.12\, M_\odot\,\rm pc^{-3}$ for these galaxies, respectively. The value for Horologium I is lower but comparable to the value of  $\approx 6\, M_\odot\,\rm pc^{-3}$ reported by \citet{safarzadeh_loeb21}. For Tucana II, on the other hand, they reported an order of magnitude higher density. Nevertheless, even the higher density values reported by these authors are typical for galaxies of $M_\star\approx 5-7\times 10^3$ according to the model predictions shown in Figure~\ref{fig:rhohalf_ms}.

\citet{Kaplinghat.etal.2019} reported a tentative correlation between central densities of the observed classical dwarf galaxies within 150 pc and pericentres of their orbits,  estimated using Gaia proper motions. \citet{Pace.etal.2022}, however, showed that this correlation significantly weakens once pericentres are estimated taking into account gravitational effects of LMC. We examined the distribution of $\rho_{\rm dm}(<150\rm\,pc)$ vs pericentre for our model galaxies and did not find any detectable correlations in any of the MW host halos in the Caterpillar suite.  

Having established that model approximately reproduces the typical $M_{\rm tot}(<\rhalf)$ and $\rhohalf$ values estimated for observed MW satellites, we now consider their evolutionary histories.

\begin{figure}
   \centering
   {\includegraphics[width=0.495\textwidth]{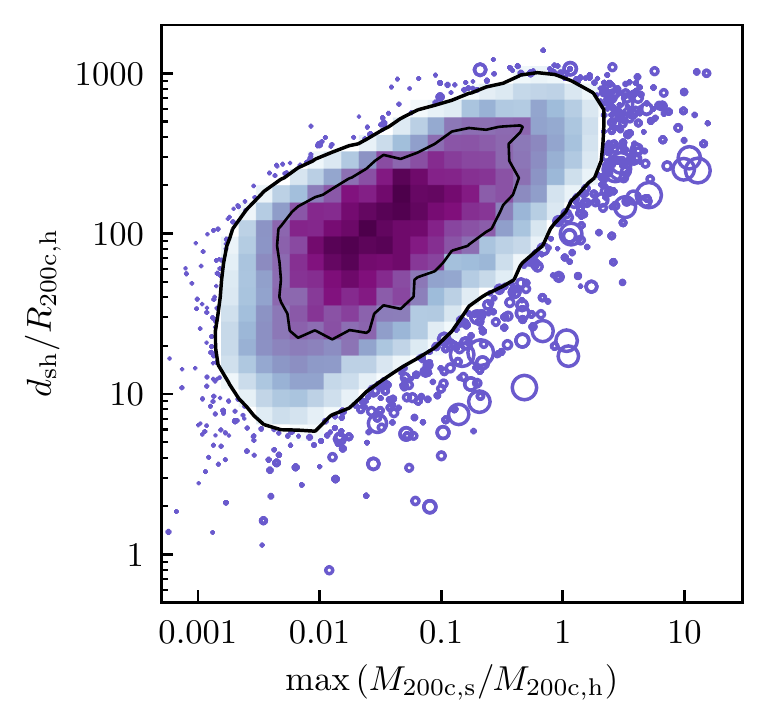}}
   \caption{Distribution of the distance between satellite progenitor divided by $R_{\rm 200c}$ of the host halo progenitor, $d_{\rm sh}/R_{\rm 200c,h}$ (shown on the $y$-axis),  at the time when the ratio of satellite and host halo progenitor masses was largest, $\max(M_{\rm 200c,s}/M_{\rm 200c,h}$ (shown on the $x$-axis). In the regions containing $68.27\%$ and $95.45\%$  of the objects (shown by black contours) distribution is shown as a 2D histogram, while outside these regions individual objects are shown as circles with the size of the circle scaling as $M_\star^{1/2.5}$, which is roughly proportional to the parent halo mass. The figure shows that although a significant fraction of satellite progenitors once (generally at $z\gtrsim 10-15$) had masses comparable to the MW host progenitor mass, the separation between satellite and MW progenitor halos was always $>5R_{\rm 200c}$ at these epochs with typical separations much larger than this lower limit.}
   \label{fig:dmrmax}
\end{figure}

%------------------------------------
\subsection{Evolution of satellite halo mass and distance from  the host halo}
\label{sec:mdevo}
%------------------------------------

Upper panels of Figure~\ref{fig:mah_revo} show evolution of halo mass of the main progenitors of the MW-sized hosts and of the subhalos that host galaxies with $z=0$ luminosities similar to those of the Horologium I, Tucana II, and Sculptor. Specifically, we select model galaxies with absolute $V$-band magnitudes within $\pm 0.5$ of the $M_V$ value of the corresponding galaxy. We show the evolution of the satellite galaxies with luminosity similar to Horologium I and Tucana II as \citet{safarzadeh_loeb21}  argued that these galaxies pose a challenge for $\Lambda$CDM. The Fornax galaxy is shown because it is an example of a massive dwarf satellite, for which the progenitor halo mass at high $z$ is particularly close to the halo mass of the MW progenitor. Overall, the evolution shown in these three panels is typical of all model satellite galaxies.  

The figure shows that the inference that satellites typically have halo mass comparable to that of the MW progenitor at high $z$ is correct. This typically occurs at $z\gtrsim 15$ for low-mass systems and at $z\gtrsim 10$ for higher-mass systems. However, as shown in the lower panels of Figure~\ref{fig:mah_revo}, when halo masses of the satellite and MW progenitors were close, these progenitors have been $\gtrsim 50$ virial radii apart. This prevented their merger during these early epochs. 
At later times, when progenitors move closer and the satellite progenitor crosses the virial radius of the MW progenitor, the halo mass ratio is $10^{-4}-10^{-3}$ and dynamical friction is inefficient. 

Figure~\ref{fig:dmrmax} shows the distance between satellite and MW progenitors (in units of the MW progenitor's $R_{\rm 200c}$)  at the time when the ratio of satellite and host halo progenitor masses was largest, $\max(M_{\rm 200c,s}/M_{\rm 200c,h})$. Remarkably, the figure shows that quite a few surviving satellites had halo masses up to a factor of $\sim 10$ larger than the MW progenitor at some early epochs. However, this must have occurred well before these objects were accreted onto MW because by definition the main progenitor of the MW must have had a larger halo mass at the time of accretion. 

Figure~\ref{fig:dmrmax} also shows that surviving subhaloes that had masses $\gtrsim 0.1$ of the MW progenitor mass were all at more than 10 virial radii apart. Conversely, Figure~\ref{fig:dmrmax}  shows 
that the lower right corner is devoid of objects. This may be because halos that may have occupied this corner of this parameter space were too close to the Milky Way progenitor and did not survive to $z=0$. 

The progenitor masses of the satellite and MW haloes have similar masses at the earliest epochs because they collapse from small-scale perturbations, which are more likely to have similar amplitude and formation redshifts due to the flattness of the $\Lambda$CDM fluctuation amplitude spectrum at small scales. However, their subsequent evolution is determined by the amplitude of the surrounding density perturbation on a larger scale or, equivalently, by the amount of mass in their immediate vicinity available for accretion by these progenitors. The MW progenitor is thus closer to the center of the large-scale density perturbation and has large amounts of mass available for accretion, while the progenitor of the satellite is located at the periphery of the perturbation and accretes at a smaller rate. 

As mass difference grows, the MW progenitor starts to dominate its environment and can stun the mass growth of neighboring halos via tidal forces.  A similar divergence of mass evolution of comparable mass halos was discussed by \citet{Bose.Deason.2023}, who found that some halos of mass $\approx 10^{11}\, M_\odot$ at $z=2$ evolve into MW-sized halos today, while others either grow much slower or merge with massive neighbors.

%------------------------------------------------------
\section{Summary and conclusions}
\label{sec:summary}
%--------------------------------

We use the \texttt{GRUMPY} galaxy formation model based on a suite of zoom-in, high-resolution, dissipationless simulations of the MW-sized haloes from the Caterpillar suite of zoom-in $\Lambda$CDM simulations to examine matter density, $\rho_{\rm tot}(<\rhalf)$, within the half-mass radius $\rhalf$ of stellar distribution and mass evolution of satellite dwarf galaxies around the Milky Way hosts. 

We compared matter densities predicted by the model to estimates of such density for 52 observed dwarf satellites of the Milky Way spanning the entire observed luminosity range using an up-to-date compilation of absolute magnitudes, half-light radii, and line-of-sight velocity dispersion measurements (Section~\ref{sec:data} and Appendix~\ref{app:sample}). Our main results and conclusions are as follows. 

\begin{itemize}
\item[(i)] We show that the model reproduces the normalization and shape of the correlation between the total mass within half-light radius, $M_{\rm tot}(<\rhalf)$ and $V$-band luminosity of observed MW satellites (see Figure~\ref{fig:mhalf_lv}), which indicates that the model forms galaxies of correct luminosity and size in haloes of a given mass. \\

\item[(ii)] We 
find that observed dwarf satellites of the Milky Way exhibit a trend of decreasing total matter density within half-light radius, $\rhohalf$, with increasing stellar mass. This trend is in general agreement with the trend predicted by our model, especially for galaxies with $M_\star>10^5\, M_\odot$.
\\
\item[(iii)] None of the observed satellites are overly dense compared to the results of our $\Lambda$CDM-based model and the scatter of their $\rhohalf$ values is comparable to model expectations.  
\\
\item[(iv)] We show that although at $z\gtrsim 10$ halo masses of many progenitors of satellites surviving to $z=0$ become comparable to or larger than the halo mass of the Milky Way progenitor, at that time the satellite and MW progenitors are separated by the distance of dozens or even hundreds of virial radii. Thus these objects do not merge during these early epochs. At the same time, because MW progenitor halo mass evolves much faster that of the satellite progenitor, by the time the latter accrete onto MW progenitor, they have typically mass ratios of $\lesssim 0.01$ (with the exception of rare major merger accretion events when and MW progenitor accretes an LMC-sized object).  
\\
\end{itemize}

Our results show that neither the densities estimated in observed Milky Way satellites nor their mass assembly histories pose a challenge to the $\Lambda$CDM model. In fact, the broad agreement between density trends with the stellar mass of the observed and model galaxies both in their form and scatter can be viewed as yet another success of the model. 
More detailed further examinations and comparisons will be warranted in the future as estimates of structural parameters and velocity dispersions of observed galaxies improve, especially for the faintest satellite galaxies. 

%----------------
% \acknowledgement
\section*{Acknowledgements}
We thank Alexander Ji and the Caterpillar collaboration for providing halo tracks of the Caterpillar simulations used in this study and Anirudh Chiti for providing us with his compilation of observational data for the Local Group dwarfs based on the updated list of values from Table 1 of \citet{Simon.2019}, which we used as a starting point for compiling the table presented in the Appendix.
This work was supported by the National Science Foundation grants AST-1714658 and AST-1911111 and NASA ATP grant 80NSSC20K0512.
Analyses presented in this paper were greatly aided by the following free software packages: {\tt NumPy} \citep{numpy_ndarray}, {\tt SciPy} \citep{scipy}, {\tt Matplotlib} \citep{matplotlib}, and \href{https://github.com/}{\tt GitHub}. We have also used the Astrophysics Data Service (\href{http://adsabs.harvard.edu/abstract_service.html}{\tt ADS}) and \href{https://arxiv.org}{\tt arXiv} preprint repository extensively during this project and the writing of the paper.

%%%%%%%%%%%%%%%%%%%%%%%%%%%%%%%%%%%%%%%%%%%%%%%%%%
\section*{Data Availability}

Halo catalogs from  the Caterpillar simulations is available at \href{https://www.caterpillarproject.org/}{\tt https://www.caterpillarproject.org/}. The \texttt{GRUMPY} model pipeline is available at \url{https://github.com/kibokov/GRUMPY}. The data used in the plots within this article are available on request to the corresponding author.

\bibliographystyle{mnras}
\bibliography{dwsat}

%%%%%%%%%%%%%%%%%%%%%%%%%%%%%%%%%%%%%%%%%%%%%%%%%%
\appendix
\section{The origin of the $M_{\rm tot}(<\rhalf)-L_V$ relation}
\label{app:lv_mhalf}

The power law relation between the total mass within the half-mass radius and $V$-band galaxy luminosity discussed in Section~\ref{sec:mhalf_lv_comp} (see eq.~\ref{eq:lv_mhalf_corr} and Figure~\ref{fig:mhalf_lv})
reflects the relation of $\rhalf$ and parent halo virial radius and the relation between luminosity and halo mass.

For example, suppose we assume 1) the approximately linear $\rhalf=\chi R_{\rm 200c}$ \citep{Kravtsov.2013}, 2) the approximately power law $L_V-M_{\rm 200c}$ relation, $L_V\propto M_{\rm 200c}^\alpha$, where $\alpha\approx 2-2.5$ \citep[e.g.,][]{Kravtsov.etal.2018,Read.Erkal.2019,Nadler.etal.2020}, 3) that dark matter dominates within $\rhalf$, and thus we can use NFW mass profile. Then $M_{\rm tot}(<\rhalf)$ can be approximated by the equation~\ref{eq:mdmnfw}, which shows that $M_{\rm tot}(<\rhalf)\propto M_{\rm 200c}$ with the factor of proportionality $f(\chi c_{\rm 200c})/f(c_{\rm 200c})$, where $c_{\rm 200c}=R_{\rm 200c}/r_s$ is halo concentration. This factor is only weakly dependent on $M_{\rm 200c}$, such that the overall relation can be accurately approximated by $M_{\rm tot}(<\rhalf)=4.25\times 10^6\,M_\odot (M_{\rm 200c}/10^8)^{0.9}$. 

Thus, using the $L_V\propto M_{\rm 200c}^\alpha$ relation we get $M_{\rm tot}(<\rhalf)\propto L_V^{0.9/\alpha}$. For $\alpha\approx 2-2.5$ the power law index is $\approx 0.36-0.45$, which is shallower than the relation we find for our model and observed galaxies. The main reason for this is that the $\rhalf-R_{\rm 200c}$ relation in the model galaxies is steeper than linear at the smallest masses and exhibits large scatter and this steepens the $M_{\rm tot}(<\rhalf)-M_{\rm 200c}$ relation to $M_{\rm tot}(<\rhalf)\propto M_{\rm 200c}^{1.2}$, which, in turn, results in $M_{\rm tot}(<\rhalf)\propto L_V^{0.55}$ relation that describes the correlation shown in Figure~\ref{fig:mhalf_lv}.

\section{Observational data}
\label{app:sample}
Table~\ref{tab:sample} shows the values of the $V$-band absolute magnitudes, $M_V$, half-light radii, $R_{1/2}$, and line-of-sight velocity dispersions of the Milky Way satellites used in this study.  The basis of the sample is  the Supplemental Table 1 in \citet{Simon.2019} review. This compilation was augmented with new measurements for existing dwarfs \citep[e.g., new measurements of velocity dispersion for the Aquarius II, Reticulum, Tucana II, etc.][]{Bruce.etal.2023,Ji.etal.2023,Chiti.etal.2023} and  for several newly discovered ultra-faint dwarfs, such as Pegasus IV \citep{Cerny.etal.2023} to the extent that we could identify such measurements. We used uniform measurements of structural properties and $M_V$ and half-light radii  from the Megacam-based study of \citet{Munoz.etal.2018} for galaxies for which these are available. Their half-light estimate for the Plummer model is used for half-light radii, because the Plummer model provides one of the best matches to the projected stellar surface density profiles of these galaxies. 

The last column in Table~\ref{tab:sample} provides references for the estimates of galaxy properties. The first reference(s) refer 
to $M_V$ and $R_{1/2}$ estimates, and the last reference to the velocity dispersion estimate, unless both were made in a single paper in which case a single reference is given. For some galaxies only upper limit on the velocity dispersion was obtained so far. We plot these as upper limits on mass and density in our analyses. For Centaurus I velocity dispersion $\sigma_{\star,\rm los}=5.5$ km/s is reported by \citet{Martinez_Vazquez.etal.2021} without uncertainties and this galaxy is shown by open star without error bar in our plots. 

Unlike other galaxies in the sample LMC and SMC are not dwarf spheroidals but of the irregular type. Velocity dispersions listed 
in the table for these galaxies are not the actual velocity dispersions, but rather values that would give the same $M(<\rhalf)$
as obtained from the estimate $M(<\rhalf)=G^{-1}v_{\rm rot,1/2}^2\rhalf$, where  $v_{\rm rot,1/2}=v_{\rm rot}(\rhalf)$ is measured rotation velocity at $\rhalf$. For the latter we use $R_{1/2}$ values listed and the table and published rotation curves for the LMC \citep[][see their Fig. 14]{Luri.etal.2021} and SMC \citep{DiTeodoro.etal.2019} to estimate $v_{\rm rot,1/2, LMC}\approx 60\pm 5$ km/s and $v_{\rm rot,1/2,SMC}\approx 33\pm 2$ km/s.  

We note that it is still debated whether some of the objects included in our sample are star clusters or galaxies (e.g., Sagittarius II, Phoenix II). We choose to do so because there is still a possibility that these may be galaxies and because velocity dispersions and half-light radii of such systems are consistent with them being dwarf galaxies. Likewise, we include 
galaxies which may be heavily influenced by tidal stripping, such as Antlia II, Crater II, Tucana III because we want to retain the full range of $\rho(<\rhalf)$ values in the observed satellites. 

\begin{table*}
	\centering
	\caption{$V$-band absolute magnitudes, $M_V$, half-light radii, $R_{1/2}$, and line-of-sight velocity dispersions of the Milky Way satellites used in this study. The last column provides references for the estimates of galaxy properties.}
	\label{tab:example_table}
	\begin{tabular}{lrrrl} 
		\hline
		Galaxy name & $M_V$ & $R_{1/2}$ & $\sigma_{\rm los}$ & References\\
                    &       &     (pc)           &        km/s  &\\
		\hline
Antlia II & $-9.86^{+0.08}_{-0.08}$ & $2541.0^{+175.0}_{-175.0}$ & $5.98^{+0.37}_{-0.36}$ & \citet{Ji.etal.2021}\\
Aquarius II & $-4.36^{+0.14}_{-0.14}$ & $159.0^{+24.0}_{-24.0}$ & $4.7^{+1.8}_{-1.2}$ & \citet{Torrealba.etal.2016b,Bruce.etal.2023}\\
Bo\"otes I & $-6.02^{+0.25}_{-0.25}$ & $191.0^{+5.0}_{-5.0}$ & $5.1^{+0.7}_{-0.8}$ & \citet{Jenkins.etal.2021}\\
Bo\"otes II & $-2.94^{+0.74}_{-0.74}$ & $38.7^{+5.1}_{-5.1}$ & $2.9^{+1.6}_{-1.2}$ & \citet{Munoz.etal.2018,Bruce.etal.2023}\\
Bo\"otes III & $-5.8^{+0.5}_{-0.5}$ & $289.0^{+100.0}_{-100.0}$ & $10.7^{+3.5}_{-3.5}$ & \citet{Correnti.etal.2009,Carlin.etal.2009,Carlin.Sand.2018}\\
Canes Venatici I & $-8.8^{+0.06}_{-0.06}$ & $452.0^{+13.0}_{-13.0}$ & $7.6^{+0.4}_{-0.4}$ & \citet{Munoz.etal.2018,Simon.Geha.2007}\\
Canes Venatici II & $-5.17^{+0.32}_{-0.32}$ & $70.7^{+11.2}_{-11.2}$ & $4.6^{+1.0}_{-1.0}$ & \citet{Munoz.etal.2018,Simon.Geha.2007}\\
Carina & $-9.43^{+0.05}_{-0.05}$ & $308.0^{+3.0}_{-3.0}$ & $6.6^{+1.2}_{-1.2}$ & \citet{Munoz.etal.2018,Walker.etal.2009}\\
Carina II & $-4.5^{+0.1}_{-0.1}$ & $91.0^{+8.0}_{-8.0}$ & $3.4^{+1.2}_{-0.8}$ & \citet{Torrealba.etal.2018,Li.etal.2018}\\
Carina III & $-2.4^{+0.2}_{-0.2}$ & $30.0^{+9.0}_{-9.0}$ & $5.6^{+4.3}_{-2.1}$ & \citet{Torrealba.etal.2018,Li.etal.2018}\\
Centaurus I & $-5.55^{+0.11}_{-0.11}$ & $79.0^{+14.0}_{-10.0}$ & $5.5$ & \citet{Mau.etal.2020,Martinez_Vazquez.etal.2021}\\
Columba I & $-4.5^{+0.17}_{-0.17}$ & $103.0^{+25.0}_{-25.0}$ & $<12.2$ & \citet{Munoz.etal.2018,Fritz.etal.2019}\\
Coma Berenices & $-4.38^{+0.25}_{-0.25}$ & $72.1^{+3.8}_{-3.8}$ & $4.6^{+0.8}_{-0.8}$ & \citet{Munoz.etal.2018,Simon.Geha.2007}\\
Crater II & $-8.2^{+0.1}_{-0.1}$ & $1066.0^{+84.0}_{-84.0}$ & $2.34^{+0.42}_{-0.30}$ & \citet{Torrealba.etal.2016}, \citet{Ji.etal.2021}\\
Draco & $-8.71^{+0.05}_{-0.05}$ & $214.0^{+2.0}_{-2.0}$ & $9.1^{+1.2}_{-1.2}$ & \citet{Munoz.etal.2018,Walker.etal.2009}\\
Draco II & $-0.8^{+0.4}_{-1.0}$ & $19.0^{+4.0}_{-2.6}$ & $<5.9$ & \citet{Longeard.etal.2018}\\
Eridanus II & $-7.21^{+0.09}_{-0.09}$ & $196.0^{+18.8}_{-18.8}$ & $6.9^{+1.2}_{-0.9}$ & \citet{Munoz.etal.2018,Li.etal.2017}\\
Fornax & $-13.46^{+0.14}_{-0.14}$ & $838.0^{3.0}_{-3.0}$ & $11.7^{+0.9}_{-0.9}$ & \citet{Munoz.etal.2018,Walker.etal.2009}\\
Grus I & $-4.1^{+0.3}_{-0.3}$ & $151.3^{+21.0}_{-31.0}$ & $2.5^{+1.3}_{-0.8}$ & \citet{Cantu.etal.2021,Chiti.etal.2022}\\
Grus II & $-3.5^{+0.3}_{-0.3}$ & $94.0^{+9.0}_{-9.0}$ & $<2.0$ & \citet{Simon.etal.2020}\\
Hercules & $-5.83^{+0.17}_{-0.17}$ & $216.0^{+17.0}_{-17.0}$ & $5.1^{+0.9}_{-0.9}$ & \citet{Munoz.etal.2018,Simon.Geha.2007}\\
Horologium I & $-3.55^{+0.56}_{-0.56}$ & $36.5^{+7.1}_{-7.1}$ & $4.9^{+2.8}_{-0.9}$ &\citet{Munoz.etal.2018,Koposov.etal.2015}\\
Horologium II & $-1.56^{+1.02}_{-1.02}$ & $44.0^{+13.8}_{-13.8}$ & $<54.6$ & \citet{Munoz.etal.2018,Fritz.etal.2019}\\
Hydra II & $-4.6^{0.37}_{-0.37}$ & $59.2^{+10.9}_{-10.9}$ & $<3.6$ & \citet{Munoz.etal.2018,Kirby.etal.2015}\\
Hydrus I & $-4.71^{+0.08}_{-0.08}$ & $53.3^{+3.6}_{-3.6}$ & $2.7^{+0.5}_{-0.4}$ & \citet{Koposov.etal.2018}\\
Leo I & $-11.78^{+0.28}_{-0.28}$ & $270.0^{+2}_{-2}$ & $9.2^{+0.4}_{-0.4}$ & \citet{Munoz.etal.2018,Walker.etal.2009}\\
Leo II & $-9.74^{+0.04}_{-0.04}$ & $171.0^{+2.0}_{-2.0}$ & $7.4^{+0.4}_{-0.4}$ & \citet{Munoz.etal.2018,Spencer.etal.2017}\\
Leo IV & $-4.99^{+0.26}_{-0.26}$ & $114.0^{+13.0}_{-13.0}$ & $3.4^{+1.3}_{-0.9}$ & \citet{Jenkins.etal.2021}\\
Leo V & $-4.29^{+0.36}_{-0.36}$ & $49.0^{+16.0}_{-16.0}$ & $2.3^{+3.2}_{-1.6}$ & \citet{Jenkins.etal.2021}\\
Leo T & $-7.6^{+0.14}_{0.14}$ & $153.0^{+16}_{-16}$ & $7.5^{+1.6}_{-1.6}$ & \citet{Munoz.etal.2018,Simon.Geha.2007}\\
LMC & $-18.1^{+0.1}_{-0.1}$ & $2697.0^{+115}_{-115}$ & $30.0^{+2.5}_{-2.5}$ & \citet{Munoz.etal.2018}, see text\\
Pegasus III & $-3.4^{+0.4}_{-0.4}$ & $53.0^{+14.0}_{-14.0}$ & $5.4^{+3.0}_{-2.5}$ & \citet{Kim.etal.2016}\\
Pegasus IV & $-4.25^{+0.2}_{-0.2}$ & $41.0^{+8.0}_{-6.0}$ & $3.3^{+1.7}_{-1.1}$ & \citet{Cerny.etal.2023}\\
Phoenix II & $-3.3^{+0.63}_{-0.63}$ & $36.0^{+12.8}_{-12.8}$ & $11.0^{+9.4}_{-5.3}$ & \citet{Munoz.etal.2018,Fritz.etal.2019}\\
Pisces II & $-4.22^{+0.38}_{-0.38}$ & $59.3^{+8.5}_{-8.5}$ & $5.4^{+3.6}_{-2.4}$ & \citet{Munoz.etal.2018,Kirby.etal.2015}\\
Reticulum II & $-3.88^{+0.38}_{-0.38}$ & $48.2^{+1.7}_{-1.7}$ & $2.97^{+0.43}_{-0.35}$ & \citet{Munoz.etal.2018,Ji.etal.2023}\\
Reticulum III & $-3.3^{+0.29}_{-0.29}$ & $64.0^{+24.0}_{-24.0}$ & $<31.2$ & \citet{Munoz.etal.2018,Fritz.etal.2019}\\
Sagittarius & $-13.5^{+0.15}_{-0.15}$ & $2662.0^{+193.0}_{-193.0}$ & $9.6^{+0.4}_{-0.4}$ & \citet{Simon.2019}\\
Sagittarius II & $-5.7^{+0.1}_{-0.1}$ & $35.5^{+1.4}_{-1.2}$ & $1.7^{+0.5}_{-0.5}$ & \citet{Longeard.etal.2020, Longeard.etal.2021}\\
Sculptor & $-10.82^{+0.14}_{-0.14}$ & $280.0^{+1.0}_{-1.0}$ & $9.2^{+1.1}_{-1.1}$ & \citet{Munoz.etal.2018,Walker.etal.2009}\\
Segue 1 & $-1.30^{+0.73}_{-0.73}$ & $24.2^{+2.8}_{-2.8}$ & $3.7^{+1.4}_{-1.1}$ & \citet{Munoz.etal.2018,Simon.etal.2011}\\
Segue 2 & $-1.86^{+0.88}_{-0.88}$ & $38.3^{+2.8}_{-2.8}$ & $<2.2$ & \citet{Munoz.etal.2018,Kirby.etal.2013segue2}\\
Sextans & $-8.72^{+0.06}_{-0.06}$ & $413.0^{+3.0}_{-3.0}$ & $7.9^{+1.3}_{-1.3}$ & \citet{Munoz.etal.2018,Walker.etal.2009}\\
SMC & $-16.8^{+0.1}_{-0.1}$ & $1106.0^{+77}_{-77}$ & $16.5^{+1}_{-1}$ & \citet{Munoz.etal.2018}, see text\\
Triangulum II & $-1.60^{+0.76}_{-0.76}$ & $17.4^{+4.3}_{-4.3}$ & $<3.4$ & \citet{Munoz.etal.2018,Buttry.etal.2022}\\
Tucana II & $-3.8^{+0.1}_{-0.1}$ & $165.0^{+27.8}_{-18.5}$ & $3.8^{+1.1}_{-0.7}$ & \citet{Koposov.etal.2015,Chiti.etal.2023} \\
Tucana III & $-1.3^{+0.2}_{-0.2}$ & $34.0^{+8.0}_{-8.0}$ & $<1.2$ & \citet{Mutlu_Pakdil.etal.2018,Simon.2019}\\
Tucana IV & $-3.0^{+0.3}_{-0.4}$ & $127.0^{+22.0}_{-16.0}$ & $4.3^{+1.7}_{-1.0}$ & \citet{Simon.etal.2020}\\
Tucana V & $-1.1^{+0.5}_{-0.6}$ & $34.0^{+11.0}_{-8.0}$ & $<7.4$ & \citet{Simon.etal.2020}\\
Ursa Major I & $-5.12^{+0.38}_{-0.38}$ & $234.0^{+10.0}_{-10.0}$ & $7.0^{+1.0}_{-1.0}$ & \citet{Munoz.etal.2018,Simon.2019}\\
Ursa Major II & $-4.25^{+0.26}_{-0.26}$ & $128.0^{+5.0}_{-5.0}$ & $5.6^{+1.4}_{-1.4}$ &\citet{Munoz.etal.2018, Simon.2019}\\
Ursa Minor & $-9.03^{+0.05}_{-0.05}$ & $405.0^{+21.0}_{-21.0}$ & $9.5^{+1.2}_{-1.2}$ & \citet{Munoz.etal.2018,Walker.etal.2009}\\
Willman 1 & $-2.53^{+0.74}_{-0.74}$ & $27.7^{+2.4}_{-2.4}$ & $4.0^{+0.8}_{-0.8}$ & \citet{Munoz.etal.2018,Willman.etal.2011}\\
\hline
\label{tab:sample}
\end{tabular}
\end{table*}

% Don't change these lines
\bsp	% typesetting comment
\label{lastpage}
\end{document}